\documentstyle[aps,prl,epsfig]{revtex}

\begin{document}

\draft

\title{Ponderomotive entanglement purification}

\author{Stefano Mancini}

\address{
INFM, Dipartimento di Fisica,
Universit\`a di Milano,
Via Celoria 16, I-20133 Milano, Italy
}

\date{\today}

\maketitle

\widetext

\begin{abstract}
It is shown that ponderomotive force can be used to 
purify entangled states. The protocol is based 
on the possibility to exploit such force for a local
quantum nondemolition measurement of the total  
excitation number of continuos variable entangled pairs.   
\end{abstract}

\pacs{PACS number(s): 03.65.Bz, 42.50.Vk, 03.67.Hk}

\section{Introduction}

Entanglement \cite{SCH} has been recognized as one of the most 
puzzling features of 
Quantum Mechanics, being also the basis of quantum information 
processing \cite{BEDI}.
However, to realize protocols like quantum key distribution or 
quantum teleportation \cite{BEN} one needs 
of maximally entangled states
which are not easy to generate (due to loss and decoherence).
Thus entanglement purification is often required to distill
maximum entangled states from several pairs 
of partial entangled states
by using local quantum operation and classical 
communications \cite{PUR}.

For qubit systems, efficient entanglement purification 
protocols have been 
found \cite{PUR}. Recently, quantum information protocols 
have been also extended 
to continuos variable systems \cite{CONT}.
In particular, in Ref. \cite{DUAN} an efficient continuos variable
entanglement purification protocol has been proposed.
Here, a physical implementation of this protocol, 
based on ponderomotive 
effect is studied.
The protocol is based on 
the possibility of a local
quantum nondemolition (QND) measurement 
of the total excitation number of continuos variables.
On the other hand the principle underlying a method of measuring 
electromagnetic energy avoiding 
photon absorption has been known for 
approximately a hundred years: measure the electromagnetic 
pressure associated with the energy 
by an experiment like that of Peotr N. Lebedev, 
and from that pressure 
infer the energy. 
Hence, we shall consider such {\it ponderomotive meter} 
\cite{BRAG} for entanglement purification.

\section{The Model}

It is well known \cite{GZ} that two-mode squeezed 
states can be generated 
as intracavity stationary state 
from the broadband squeezed light 
provided by a Nondegenerate Optical Parametric 
Amplifier (NOPA) operating below threshold.
Let us now suppose that the two modes characterize 
two different cavities.
The latter are also placed in two distant sides, 
say $A$ and $B$, respectively.
Then, the steady state of the cavity modes $a_1$ (in the A side)
and $b_1$ (in the B side) can be written as
$S(r)|0\rangle_{a_1}|0\rangle_{b_1}$ where
$S(r)=\exp[r(a_1^{\dag}b_1^{\dag}-a_1b_1)]$ is the 
squeezing operator with the squeezing parameter $r$ \cite{CAVES}.
Now, let us consider another pair of modes, 
say $a_2$ (in the A side) and $b_2$ (in the B side),
in the same state, i.e. $S(r)|0\rangle_{a_2}|0\rangle_{b_2}$.
Furthermore, 
suppose that the $a_i$ ($b_i$) modes $i=1,2$, belong
to the same cavity in the A (B) side. For instance, they 
have the same frequency $\omega_a$ ($\omega_b$),
but they distinguish for their polarization.
Then, the state of the whole system is
$|\psi\rangle_1\otimes|\psi\rangle_2$,
where $|\psi\rangle_i$ ($i=1,2$) is a two-mode squeezed state         
\begin{equation}\label{Psii}
|\psi\rangle_i=\sqrt{1-\lambda^2}\sum_{n=0}^{\infty}
\lambda^n |n\rangle_{a_i} \otimes |n\rangle_{b_i}\,,
\end{equation}
with $\lambda=\tanh r$.
The state $|\psi\rangle_1\otimes|\psi\rangle_2$
represents two pairs of partial entangled states.
 
Following the avenue sketched in Ref.\cite{DUAN} 
we choose to make a QND measurement of the 
total number of photons in the A side.
To this end we exploit a ponderomotive
meter \cite{BRAG}.
A suspended mirror provides an intuitive realization
of such meter \cite{JAC}, however, a more realistic apparatus
can be realized by using for example
a cavity made by a piezoelectric crystal 
coated on both sides \cite{FABRE}. 
For a piezoelectric crystal, the length variation 
due to radiation pressure can be 
measured by a suitable electric circuit.

A simple model to describe the coupling of a ponderomotive meter 
with the A modes, 
is obtained as straightforward generalization 
of that presented in Refs.\cite{MMT,BOSE}.
Let $c$ the oscillatory mode characterizing the meter and $\Omega$
its frequency,
then the Hamiltonian (in natural unit) 
governing the measurement dynamics is 
\begin{equation}\label{H}
H=\omega_a\sum_{i=1}^2 a^{\dag}_i a_i
+\omega_b\sum_{i=1}^2 b^{\dag}_i b_i
+\Omega c^{\dag}c 
-g\left[\sum_{i=1}^2 a^{\dag}_i a_i\right] 
\left(c+c^{\dag}\right)\,,
\end{equation}
with $g$ the coupling constant between signal and meter.
The interaction part of the Hamiltonian 
(fourth term in Eq.(\ref{H})),
simply represents the effect of the radiation pressure force which
causes the instantaneous displacement $x$ of the meter \cite{MT}.
Here, $x=(c+c^{\dag})/\sqrt{2}$,
$p=-i(c-c^{\dag})/\sqrt{2}$
are the meter position and momentum quadratures respectively.

On the other hand, the meter would be connected with a readout 
apparatus, and this can be modeled as a heath bath \cite{ZUR,WAL}.
In particular, a meter undergoing quantum Brownian motion 
can be described by the following master equation \cite{GZ}
\begin{equation}\label{master}
\dot{\rho}=-i\left[H,\rho\right]
-i\frac{\gamma}{2}\left[x,\left\{p,\rho\right\}\right]
-\frac{\gamma}{\beta}\left[x,\left[x,\rho\right]\right]\,,
\end{equation}
where $\gamma$ is the damping rate and $\beta$ the
parameter proportional to the inverse of the temperature.
By assuming the meter initially in the vacuum, 
the initial state of the whole system is
\begin{equation}\label{rho0}
\rho_0=|\psi\rangle_1\langle\psi|\otimes
|\psi\rangle_2\langle\psi|\otimes
|0\rangle_m\langle 0|\,.
\end{equation}
Since, from Eq.(\ref{H}), $H$ commutes with the total number 
operator in the A side, $N=\sum_{i=1}^2 a_i^{\dag} a_i$, 
one can solve Eq.(\ref{master}) by introducing the following
characteristic function
\begin{equation}\label{chi}
\chi_{N,M}(\mu,\mu^*,t)={\rm Tr}_m
\left\{\rho_{N,M}(t)e^{\mu c^{\dag}}e^{-\mu^* c}\right\}\,,
\end{equation}
where $\rho_{N,M}$ indicate the density matrix elements
of radiation fields built up with `ket' and `bra' having 
total number of photons $N$ and $M$ respectively.
Then, Eq.(\ref{master}) can be transformed 
in a partial differential 
equation for the characteristic function (\ref{chi}). 
The latter, in turn, due to its 
linearity allows a simple Gaussian solution.
Without going down in the details of calculations, 
it is possible to see that, in the limit $t\gg\gamma^{-1}$, 
the following steady state is reached  
\begin{eqnarray}\label{rhoss}
\rho_{ss}=\sum_{N=0}^{\infty}\sum_{n_1=0}^{N}\sum_{m_1=0}^{N}
(1-\lambda^2)^2\lambda^{2N}\,
&\,&|n_1\rangle_{a_1}\langle m_1|\otimes
|N-n_1\rangle_{a_2}\langle N-m_1|
\nonumber\\
&\otimes&|n_1\rangle_{b_1}\langle m_1|\otimes
|N-n_1\rangle_{b_2}\langle N-m_1|
\nonumber\\
&\otimes&
\left[D(-\kappa N)\rho_{th}^{(m)}D^{\dag}(-\kappa N)\right]\,,
\end{eqnarray}
where $D$ denotes the displacement operator \cite{GLA},
and $\kappa=g/\Omega$.
Moreover,
\begin{equation}\label{rhoth}
\rho_{th}^{(m)}=(1-e^{-\beta})\sum_{n=0}^{\infty}
e^{-n\beta} \, |n\rangle_m\langle n|\,,
\end{equation}
is the thermal state of the meter.
Practically, the dynamics leading to the state (\ref{rhoss})
corresponds to the `collapse' of wavefunction due to 
the measurement process \cite{NEU}. Once
$\gamma$ becomes large (strong
meter-environment interaction),
the model provides an idealized limit of 
real measurement process. Notice that 
by using the quantum Brownian motion master equation 
no Rotating Wave Approximation has been 
invoked, differently from Ref.\cite{WAL}
where the quantum optical master equation was used 
in an analog measurement model.

The radiation fields state corresponding 
to the meter outcome $x$ is
then obtained by projecting (\ref{rhoss}) 
onto the meter eigenstate
$|x\rangle_m$ \cite{NEU}, that is 
\begin{eqnarray}\label{rhorad}
\rho^{(r)}&=&{\cal N}\sum_{N=0}^{\infty}
\sum_{n_1=0}^{N}\sum_{m_1=0}^{N}
(1-\lambda^2)^2\lambda^{2N}\,
{}_m\langle x|\left[D(-\kappa N)\rho_{th}^{(m)}
D^{\dag}(-\kappa N)
\right]|x\rangle_m
\nonumber\\
&&|n_1\rangle_{a_1}\langle m_1|\otimes
|N-n_1\rangle_{a_2}\langle N-m_1|
\otimes|n_1\rangle_{b_1}\langle m_1|\otimes
|N-n_1\rangle_{b_2}\langle N-m_1|\,,
\end{eqnarray}
where the normalization constant is given by
\begin{equation}\label{calN}
{\cal N}^{-1}=\sum_{N=0}^{\infty}
{}_m\langle x|\left[D(-\kappa N)\rho_{th}^{(m)}D^{\dag}(-\kappa N)
\right]|x\rangle_m\,,
\end{equation}
and
\begin{equation}\label{xDal}
{}_m\langle x|\left[D(-\kappa N)\rho_{th}^{(m)}D^{\dag}(-\kappa N)
\right]|x\rangle_m=
\frac{1-e^{-\beta}}{\sqrt{\pi}}
\sum_{n=0}^{\infty}\frac{e^{-\beta}}{2^n n!}
\exp\left[-\left(x+\sqrt{2}\kappa N\right)^2\right]
H_n^2\left[x+\sqrt{2}\kappa N\right]\,,
\end{equation}
with $H_n(.)$ the Hermite polynomials.

\begin{figure}[t]
\centerline{\epsfig{figure=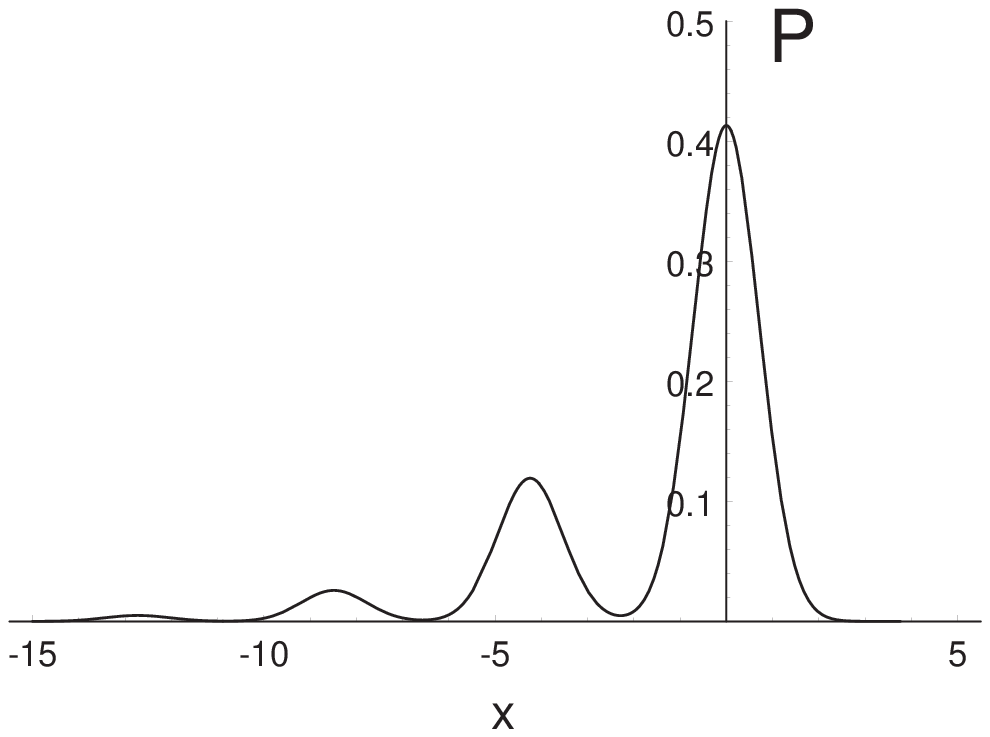,width=3.5in}}
\caption{\widetext 
Meter position probability distribution $P(x)$.
Here $\kappa=3$, $\beta=\infty$ and $r=0.4$.
}
\label{pepfig1}
\end{figure}

The normalization constant (\ref{calN}) 
depends on the value of the outcome $x$,
and gives the meter position distribution $P(x)$,
that is $P(x)\equiv {\cal N}^{-1}$.
Let us now assume for the sake of 						   
simplicity $\beta=\infty$, i.e.					   
the meter goes in a coherent state 
of amplitude $-\kappa N$ \cite{ME}.	
Then, Fig.(\ref{pepfig1}) shows 
a typical distribution of outcomes.
Different peaks correspond to different values of $N$,
from the right to the left $N=0,1\ldots$. 
Notice that in order to distinguish all possible values of $N$
one should have the minimum amount of displacement 
(corresponding to $\kappa$) greater than the width of a
coherent state, i.e. $1/2$.
Therefore, the measurement efficiency 
increases monotonically with the signal-meter 
correlation.

\section{Entanglement Quantification}

At the end of the measurement process
the partial density operator can be obtained by 
tracing Eq.(\ref{rhorad}) over the A side 
\begin{eqnarray}\label{rhoB}
\rho_B^{(r)}(x)&=&{\rm Tr}_A\left\{\rho^{(r)}(x)\right\}
\nonumber\\
&=&{\cal N}(1-\lambda^2)^2\sum_{N=0}^{\infty}\sum_{n_1=0}^{N}
\lambda^{2N}
{}_m\langle x|\left[D(-\kappa N)\rho_{th}^{(m)}D^{\dag}(-\kappa N)
\right]|x\rangle_m\;
|n_1\rangle_{b_1}\langle n_1| \otimes
|N-n_1\rangle_{b_2}\langle N-n_1|
\,.
\end{eqnarray}
One can easily see from the above expression (\ref{rhoB})
that the state $|n_1\rangle_{b_1}\,|n_2\rangle_{b_2}$
is an eigenstate of the reduced density operator
\begin{equation}\label{eigen}
\rho_B^{(r)}(x)\left(|n_1\rangle_{b_1}|n_2\rangle_{b_2}\right)
=\Lambda_{n_1,n_2}(x) \left(|n_1\rangle_{b_1}
|n_2\rangle_{b_2}\right)\,,
\end{equation}
whose eigenvalues are
\begin{equation}\label{Lambda}
\Lambda_{n_1,n_2}(x) = {\cal N}
(1-\lambda^2)^2\lambda^{2(n_1+n_2)} 
{}_m\langle x|\left[D(-\kappa N)\rho_{th}^{(m)}D^{\dag}(-\kappa N)
\right]|x\rangle_m\,.
\end{equation}

Now, the measure of entanglement for a pure
bipartite system is just the Von Neumann entropy 
of either partial density operator \cite{BHPS}.
Hence, it can be written as \cite{BOSE99}
\begin{equation}\label{calE}
{\cal E}(x)=-\sum_{n_1,n_2=0}^{\infty}
\Lambda_{n_1,n_2}(x)
\left[ \log_2 \Lambda_{n_1,n_2}(x) \right]\,,
\end{equation}
where the dependence from $x$ indicates the dependence from
the outcome of measurement.

Furthermore, the quantity 
\begin{equation}\label{Ga}
\Gamma(x) = \frac{{\cal E}(x)}{{\cal E}_0}\,,
\end{equation}
defines the entanglement increase ratio,
once ${\cal E}_0$ identifies 
the initial amount of entanglement. 
Of course, if $\Gamma(x) > 1$, the state after 
the measurement is more 
entangled with respect to the initial one.
Prior the measurement we have $\Lambda_{n_1,n_2}=(1-\lambda^2)^2
\lambda^{2(n_1+n_2)}$ and therefore
\begin{equation}\label{calE0}
{\cal E}_0=2\left[
\cosh^2r\log_2\left(\cosh^2r\right)
-\sinh^2r\log_2\left(\sinh^2r\right)\right]\,,
\end{equation}
which is obviously not depending on $x$. Furthermore, it results 
equal to the entanglement of a single pair times the number 
of pairs.

\begin{figure}[t]
\centerline{\epsfig{figure=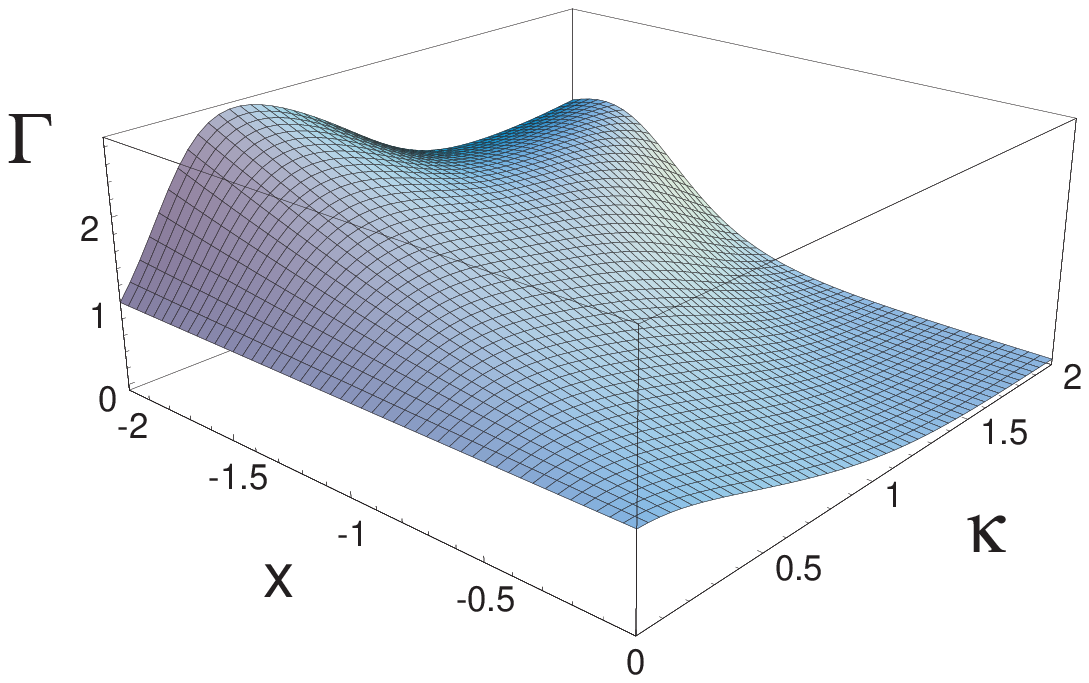,width=3.5in}}
\caption{\widetext 
The entanglement purification ratio $\Gamma$ as function
of position quadrature $x$ and coupling constant $\kappa$.
The parameters are $\beta=\infty$ and $r=0.3$.
}
\label{pepfig2}
\end{figure}

\begin{figure}[t]
\centerline{\epsfig{figure=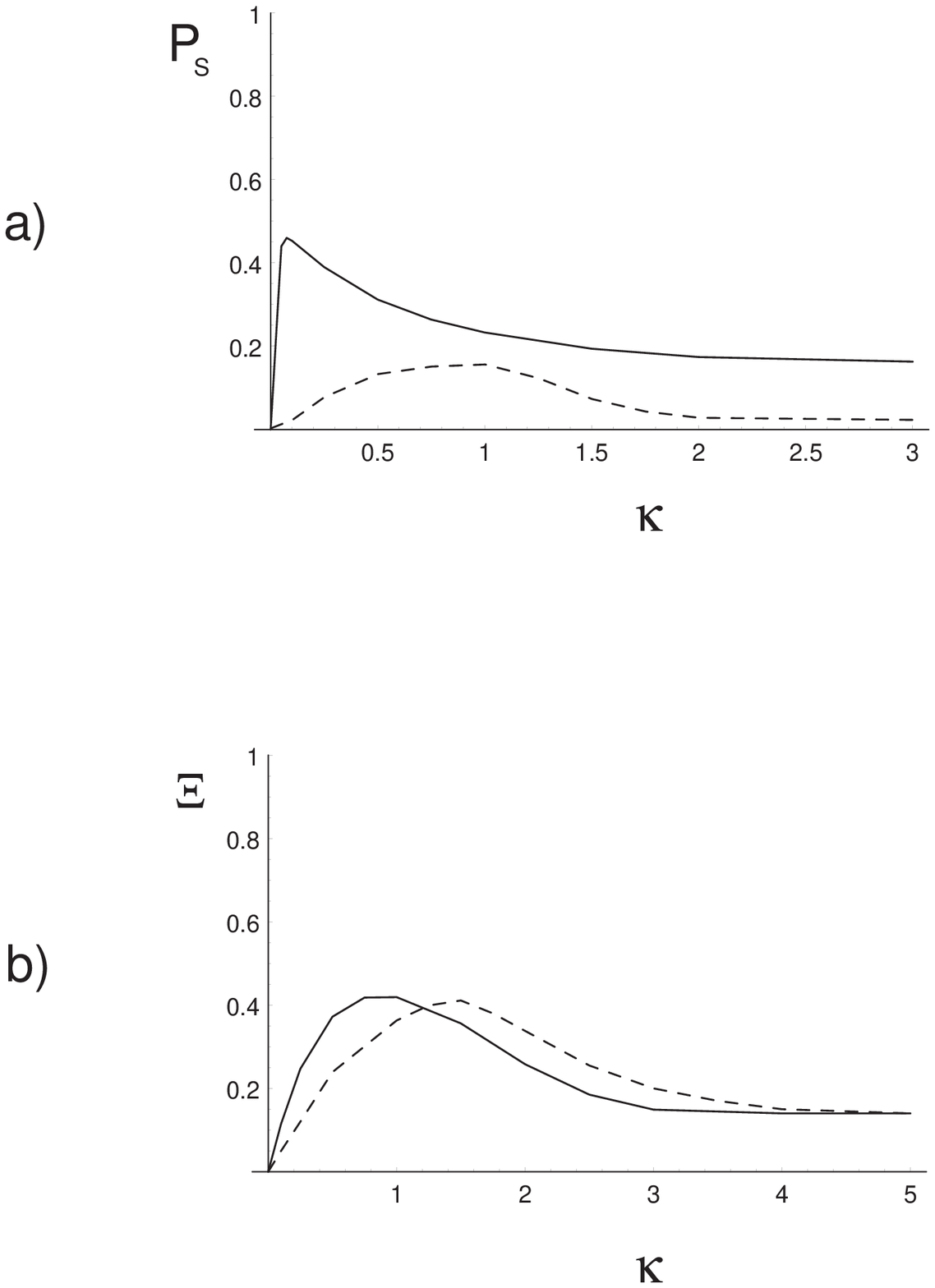,width=3.5in}}
\caption{\widetext 
The success probability $P_S$ is shown in figure a)
vs the coupling constant $\kappa$.
The solid line is for $\epsilon=1$ and the dashed line 
for $\epsilon=1.5$.
The efficiency $\Xi$ of the purification protocol 
is shown in figure b) vs the coupling constant $\kappa$
for $\epsilon=1$.
The solid line is for $\beta=\infty$ and the dashed line 
for $\beta=1$.
In both figures a) and b) $r=0.3$.
}
\label{pepfig3}
\end{figure}

In Fig. (\ref{pepfig2}) the entanglement increasing
ratio $\Gamma$ is shown as 
function of the measured value of position
quadrature $x$
and coupling constant $\kappa$.
We may see that for $\kappa=0$ the entanglement remain the same 
since the meter is practically 
uncorrelated with the signal.
On the other hand, the result $x=0$ 
(corresponding to $N=0$) never 
increases the entanglement.
For $\kappa\ne 0$, $\Gamma$ grows up 
approximately as $x$ diminishes;
however, one should take into account that the probability to 
get large negative values of $x$ decreases, as can be seen from 
Fig.(\ref{pepfig1}).
Then, it is useful to introduce the probability of success of the 
protocol
\begin{equation}\label{PS}
P_S=\int_{\Gamma>\epsilon}\,dx\, P(x)\,,
\end{equation}
where integration is performed 
over the set $\{x\in {\bf R}|\Gamma(x) > \epsilon\}$,
with $\epsilon$ defining the lower limit for the success
($\epsilon\ge 1$).
The quantity $P_S$ is plotted in Fig.(\ref{pepfig3}.a) 
for $\epsilon=1$ (solid line) and for $\epsilon=1.5$
(dashed line).
One can see that an higher increment of the degree of entanglement 
is achieved by increasing the value of $\kappa$, but with a 
lower probability.

It is also instructive to consider the efficiency of the protocol.
It can be defined as
\begin{equation}\label{Xi}
\Xi=1-\frac{1}{\Upsilon}\,,
\end{equation}
where 
\begin{equation}\label{Ups}
\Upsilon=\frac{\int_{\Gamma>\epsilon}\, dx\, \Gamma(x) P(x)}
{\int_{\Gamma>\epsilon}\, dx\, P(x)}\,,
\end{equation}
represents the average of the entanglement increase ratio weighted
on the probability of success.   
It always results $\Upsilon\ge 1$ and $\Xi\le 1$.
Fig.(\ref{pepfig3}.b) (solid line) 
shows the efficiency of the protocol.
Contrarily to what one can expect
for a better resolution of the number of photons,
the efficiency does not
increase monotonically with $\kappa$.
Instead, there exist a particular value of $\kappa$ that maximizes
the efficiency, and it is slightly greater than the width of the 
final meter state.
In fact, for such value of $\kappa$,
a measurement giving a result, say $x$, 
different from zero cannot be identified 
with a particular value of $N$,
but it practically corresponds to any $N\ne 0$, thus improving the
efficiency of the purification process.
As matter of fact we have seen in Fig.(\ref{pepfig2})
that $\Gamma$ exceeds $1$ for $N\ne 0$ (essentially $x< 0$).
However, this limit, hence the optimal value of $\kappa$, 
also depends on the initial amount of entanglement, 
i.e. on the squeezing parameter $r$. 

Fig.(\ref{pepfig3}.b) (dashed line) 
shows the thermal effects on the
efficiency of the purification protocol. 
In such a case the distribution $P$ becomes broader, 
hence, the protocol
results less efficient for $\kappa$ maximizing $\Xi$
at $\beta=\infty$.
However, the effect of temperature is somehow equivalent
to decreasing the coupling constant $\kappa$,
in fact for $\kappa\to 0$ it should be $\Gamma(x)\to 1$.
Hence the thermal effect can be hindered
by increasing $\kappa$. The extension of previous arguments
yields $\kappa$ slightly greater than 
$\beta^{-1}$ as criteria for efficient protocol.
Notice that smaller values of $\beta$ could 
be considered obtaining the same behavior \cite{ME}. 

Finally, the developed theory can be extended
to consider more than two radiation field pairs per side.
Consequently, the efficiency of the protocol can be improved.
Furthermore, in this scheme,
all the modes in one side can be considered belonging to 
the same cavity, at least 
in the side where the measurement is performed (A).
However, in that case  the coupling constant becomes mode depending
since we have to account for the mode
spacing.

\section{Conclusions}

In conclusion, the role of a ponderomotive meter in 
entanglement purification process is studied.
It is found that there exist optimal values of the
signal-meter coupling constant, contrarily from 
what one can expect by a simple analysis 
on indirect measurement.
The purification 
process results sensible to the thermal 
effects introduced by the meter,
however, appropriate signal-meter coupling
reduces such detrimental effects.
It is worth noting that a wide range of $\kappa$ values  
are accessible for movable mirrors
with the present technology \cite{EXP}.
Otherwise, microfabricated cantilevers \cite{ROU}
could be also employed.
Furthermore, a large number of pairs involved in
the process renders it more efficient.

Since radiation pressure allows to couple
macroscopic objects (like mirrors) with microscopic
systems (like cavity modes), it provides
a unique bridge between quantum and classical
worlds.
Hence, optomechanical systems are worth studying,
also for quantum information processing.

Finally, ponderomotive effects come out  
also in the dispersive regime of 
atom-field interaction \cite{ATOMS},
where the atomic momentum becomes sensible to the 
photon statistics,
and a scheme analogous to that described
can be developed with atoms crossing a cavity.
In such a case the thermal effects 
should be less significant.

\section*{Acknowledgements}
The author gratefully acknowledge financial support from
Universit\`a di Camerino, Italy, under the Project
`Giovani Ricercatori'.

\begin{thebibliography}{99}
\bibitem{ME}
To be rigorous, the master equation (\ref{master}) is only
valid for $\beta \ll 1$, however, the opposite limit
has been chosen to easily describe the qualitative behavior 
of the system.
\end{thebibliography}
\end{document}